\documentclass[aps,prl,twocolumn,notitlepage,superscriptaddress,longbibliography]{revtex4-1}
\usepackage[colorlinks=true, citecolor=magenta, linkcolor=blue, urlcolor=blue]{hyperref}
\usepackage{graphicx}
\usepackage{amsmath}
\usepackage{amssymb}
\usepackage{amsfonts}
\usepackage{hyperref}
\usepackage{mathtools}
\usepackage{xcolor}
\usepackage{verbatim}
\usepackage{ulem}

\usepackage{tikz}
\usetikzlibrary{shapes.geometric, arrows}
\usetikzlibrary{decorations.markings}
\usetikzlibrary{decorations.pathmorphing}
\usetikzlibrary{decorations.pathreplacing}
\usetikzlibrary{shapes.symbols}

\newcommand{\br}{{\bf r}}

\newcommand{\bq}{{\bf q}}

\newcommand{\bs}{{\bf s}}
\newcommand{\bS}{{\bf S}}

\allowbreak

\begin{document}
 
\title{Magnon-mediated topological superconductivity in a quantum wire}
\author{Florinda Vi\~nas Bostr\"om}
\email{florinda.vinas\_bostrom@ftf.lth.se}
\affiliation{Institut f\"ur Mathematische Physik, Technische Universit\"at Braunschweig, D-38106 Braunschweig, Germany}
\affiliation{Division of Solid State Physics and NanoLund, Lund University, Box 118, S-221 00 Lund, Sweden}
\author{Emil Vi\~nas Bostr\"om}
\email{emil.bostrom@mpsd.mpg.de}
\affiliation{Nano-Bio Spectroscopy Group, Departamento de F\'isica de Materiales, Universidad del Pa\'is Vasco, 20018 San Sebastian, Spain}
\affiliation{Max Planck Institute for the Structure and Dynamics of Matter, Luruper Chaussee 149, 22761 Hamburg, Germany}
\date{\today}

\begin{abstract}
 Many emergent phases of matter stem from the intertwined dynamics of quasi-particles. Here we show that a topological superconducing phase emerges as the result of interactions between electrons and magnons in a quantum wire and a helical magnet. The magnon-mediated interaction favors triplet superconductivity over a large magnetic phase space region, and stabilizes topological superconductivity over an extended region of chemical potentials. The superconducting gap depends exponentially on the spin-electron coupling, allowing it to be enhanced through material engineering techniques.
\end{abstract}

\maketitle


The intertwined dynamics of collective excitations underlies many diverse phenomena observed in condensed matter systems. A prominent example where such interactions play a key role is in the effective phonon-mediated attraction between electrons, responsible for the superconducting instability of simple metals at low temperatures~\cite{Bardeen1957}. Other emergent phases driven by quasi-particle interactions include the spontaneous electric polarization of ferroelectric and multiferroic materials in response to lattice distortions~\cite{Eerenstein2006,Li2019}, the binding between electrons and holes and their possible subsequent condensation into an excitonic insulator~\cite{Werdehausen2018,Mazza2020}, and the stabilization of fractional quantum Hall states through electron-electron interactions~\cite{Laughlin1983}.

Many emergent phases themselves give rise to quasi-particles with potentially unconventional properties. In particular, the excitations of topologically ordered systems~\cite{Wen1990,Kitaev2003,Nayak2008} such as fractional quantum Hall systems, quantum spin liquids, and topological superconductors~\cite{Arovas1984,Kitaev2006,Savary2016,Bartolomei2020,Nakamura2020}, termed anyons~\cite{Leinaas1977,Wilczek1982}, have garnered much interest due to their promise in realizing fault-tolerant quantum computing. In particular, the sub-class known as non-abelian anyons allow for information to be non-locally encoded and processed in the braiding patterns of the anyon world lines~\cite{Kitaev2003,Nayak2008}. In topological superconductors, non-abelian anyons may appear in vortex cores of chiral two-dimensional superconductors~\cite{Fu2008,Nakosai2013,Sun2016,Menard2017,PalacioMorales2019,Wang2020,Kezilebieke2020}, or at the ends of one-dimensional (1-D) quantum wires~\cite{Flensberg2010,Oreg2010,Lutchyn2010,Lutchyn2018,Law2009,Stanescu2011,Alicea2011,Mourik2012,Das2012,Flinck2013,Albrecht2016,Deng2016,Nichele2017,Zhang2019,Prada2020,Microsoft2023}. However, to stabilize a topological superconducting phase, the pairing needs to be mediated by quasi-particles carrying an intrinsic spin structure that favors spin triplet over spin singlet pairing.


\begin{figure}[ht]
 \includegraphics[width=\columnwidth]{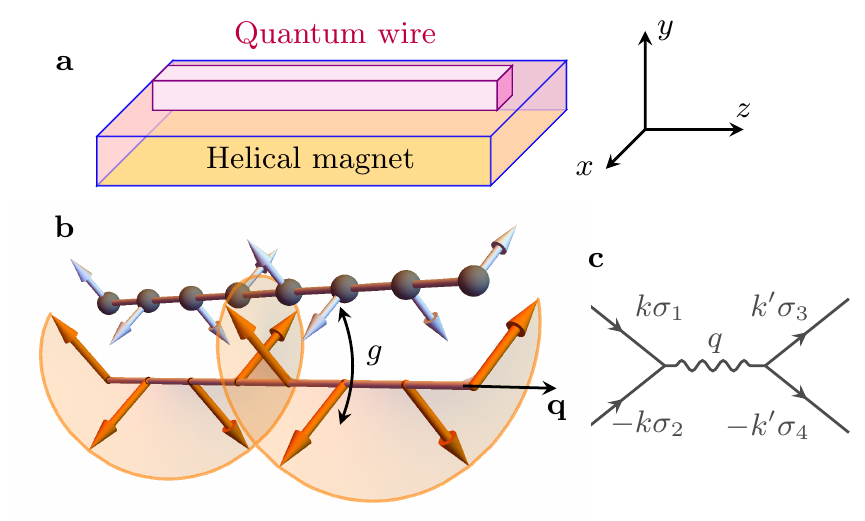}
 \caption{{Principles of magnon-mediated superconductivity.} {\bf a,} Experimental setup with a quantum wire in proximity to a helical magnet. {\bf b,} The helical magnetic order (orange arrows) induces an effective spin-orbit interaction and Zeeman splitting of the electronic bands via the spin-electron coupling $g$. Magnon fluctuations around the equilibrium magnetic order provide an effective attractive interaction among the electrons (blue arrows). {\bf c,} Due to the non-collinear magnetic structure, magnon fluctuations mediate scattering between electrons with arbitrary spin projections $\sigma_i$.}
 \label{fig:setup}
\end{figure}


As a prominent example, magnon-mediated topological superconductivity has recently been proposed for two-dimensional systems with non-collinear magnetic order, such as skyrmion crystals and helical magnets in proximity to a two-dimensional normal metal surface~\cite{Maeland2022,Maeland2023,Maeland2023b}. However, since the majority of experimental work on topological superconductivity is concerned with one-dimensional structures, identifying criteria for realizing magnon-mediated topological superconductivity in one dimension is of key importance. Indeed, both the superconducting and topological phenomenology is qualitatively different in one and two dimensions, stemming in large part from the different symmetry requirements necessary to realize a non-trivial topology in different dimensions~\cite{Altland1997,Kitaev2001,Chiu2016}.  In particular, to stabilize topological superconductivity in 2-D it is necessary to obtain a time-reversal symmetric and fully gapped superconducting state, while in 1-D systems a topological superconducting phase is realized by breaking time-reversal symmetry (thereby realizing an effective single-band regime), while simultaneously stabilizing triplet pairing. As will be demonstrated below, both these conditions are satisfied by the dynamical coupling between itinerant electrons and the magnons of a helical magnet, away from the collinear ferromagnetic and antiferromagnetic limits.


Specifically, we here investigate superconductivity resulting from the coupling between a quantum wire and a helical magnet (see Fig.~\ref{fig:setup}). The magnons of the helical magnet mediate an effective attraction between electrons of arbitrary spin projection, thereby stabilizing unconventional triplet superconductivity over a large region of phase space. The non-collinear magnetic order induces an effective spin-orbit coupling (SOC) and Zeeman field among the electrons, that allows to realize an effective single-band regime over a finite range of chemical potentials.  Within the single-band regime, the system enters a topological phase, with unpaired Majorana bound states at each end of the wire. Crucially, both the size of the effective single-band regime and the superconducting gap are increasing functions of the spin-electron coupling $g$. Our proposal thereby identifies quantum wires in proximity to helical magnets as a promising platform to realize topological superconductivity, without the need to proximitize the wire to a conventional superconductor. It also showcases the power of utilizing quasi-particle interactions to realize unconventional forms of matter.


To investigate magnon-mediated superconductivity in a quantum wire, we consider an interacting system of spins and electrons described by the lattice Hamiltonian
\begin{align}\label{eq:hamiltonian}
 H = &-t\sum_{\langle ij\rangle\sigma} \hat{c}_{i\sigma}^\dagger \hat{c}_{j\sigma} - J \sum_{\langle ij\rangle} \hat{\bS}_i \cdot \hat{\bS}_j \\
 &- \sum_{\langle ij\rangle} {\bf D}_{ij} \cdot (\hat{\bS}_i \times \hat{\bS}_j) - g \sum_i \hat{\bs}_i \cdot \hat{\bS}_i. \nonumber
\end{align}
Here $\hat{c}_{i\sigma}$ destroys an electron at site $i$ and of spin projection $\sigma$, and $\hat{\bS}_i$ is the spin operator at site $i$ for a spin of magnitude $S$. The parameters $t$, $J$, and ${\bf D}$ respectively determine the nearest-neighbor electronic hopping amplitude, the exchange interaction, and the antisymmetric Dzyaloshinskii-Moriya interaction (DMI). The spins and electrons interact via a local spin-electron coupling of strength $g$, and the electronic spin operator is defined by $\hat{\bs}_i = \sum_{\sigma\sigma'} \hat{c}_{i\sigma}^\dagger \boldsymbol\tau_{\sigma\sigma'} \hat{c}_{i\sigma'}$, with $\boldsymbol\tau$ the Pauli matrix vector.

The competition between exchange and DMI stabilizes a helical magnetic order, characterized by a propagation vector $\bq$ and a vector $\hat{\bf n}$ defining the plane of polarization. Specifically, the magnitude of the spiral momentum is given by $\tan q = D/J$, while the directions of the momentum and polarization vectors are determined by the normalized DMI vector $\hat{\bf D}$. Here we consider a DMI such that the spiral momentum ${\bf q}$ lies along the wire axis $\hat{\bf z}$, and with $\hat{\bf n}$ parallel to $\bq$ (see Fig.~\ref{fig:setup}). The equilibrium spin texture is then a spin helix, and can be written as ${\bf S}_i = \cos(\bq \cdot \br_i) {\bf e}_1 + \sin(\bq \cdot \br_i) {\bf e}_2$, where the vectors ${\bf e}_\alpha$ and $\hat{\bf n}$ define a right-handed orthonormal system~\cite{Supplemental}. 

In the following the magnetic texture is assumed to be commensurate with the electronic lattice, such that the spin spiral is periodic over a distance $La$, where $L$ is an integer and $a$ is the electronic lattice parameter. To describe fluctuations around the equilibrium order, the spin operators $\hat{\bf S}_i$ are expressed in terms of a set of bosonic operators by performing a Holstein-Primakoff expansion around the local spin axis ${\bf S}_i$~\cite{Supplemental}. This results in a diagonal magnon Hamiltonian $H_m = \sum_{np} \Omega_{np} \alpha_{np}^\dagger \alpha_{np}$, where $\Omega_{np}$ is the energy of a magnon with momentum $p$ in band $n$. The momentum runs over the magnetic Brillouin zone $[-\pi/La, \pi/La]$, and the number of magnon bands is $L$.

\begin{figure*}
 \includegraphics[width=\textwidth]{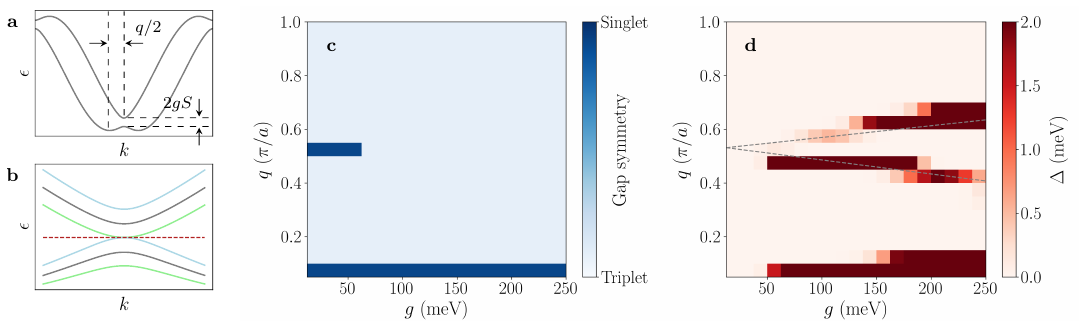}
 \caption{{Parameter dependence of magnon-mediated superconductivity.} {\bf a,} Electronic band structure for a spin-electron coupling $g = 150$~meV and a spiral momentum $q = \pi/3a$. {\bf b,} Electronic band structure around $k = 0$ for a spin-electron coupling $g = 150$~meV and spiral momenta $q = 0.44 \pi/a$ (green), $q = 0.5 \pi/a$ (gray) and $q = 0.57 \pi/a$ (blue). The red dashed line indicates the chemical potential $\mu = -1.4$~eV. {\bf c,} Symmetry of the dominant superconducting gap as a function of spin-electron coupling $g$ and spiral momentum $q$, for a chemical potential $\mu = -1.4$~eV. {\bf d,} Magnitude of the dominant superconducting gap as a function of spin-electron coupling $g$ and spiral momentum $q$, for a chemical potential $\mu = -1.4$~eV. The gray dashed lines indicate the positions of the van Hove singularities. In all panels the spin length is $S = 1$, the electronic hopping is $t = 1$~eV, and the exchange interaction is $J = 10$~meV.}
 \label{fig:parameters}
\end{figure*}

There are two main effects of the spin spiral on the electronic structure: First, the coupling to the equilibrium magnetic structure $\bS_i$ induces an effective SOC and Zeeman splitting of the electronic bands (see Fig.~\ref{fig:parameters}a). Second, the coupling to magnon fluctuations around the helical configuration generates an effective attractive interaction among the electrons, which ultimately leads to superconductivity. The effect of the static spin spiral can be exactly accounted for, by diagonalizing the electronic sub-system in presence of the spiral $\bS_i$~\cite{Bostrom2021,Supplemental}, and results in the electron Hamiltonian $H_e = \sum_{k\tau} \epsilon_{k\tau} d_{k\tau}^\dagger d_{k\tau}$. Here $\tau$ denotes the electronic bands, and the momentum runs over the electronic Brillouin zone $[-\pi/a, \pi/a]$. The spiral shifts the minima of the originally spin-degenerate bands to $\pm q/2$ (Fig.~\ref{fig:parameters}a), thereby inducing an effective Rashba SOC. This shift is independent of the value of $g$, and is set for any finite $g$ by the momentum of the spin spiral. In addition, the coupling opens a gap of size $\Delta_b = 2gS$ at $k = 0$ and $k = \pm \pi/a$, acting like an effective Zeeman field. Together these effects realize an effective single-band regime for chemical potentials inside the gap. The static effects of our model have been considered within the context of so-called Yu-Shiba-Rusinov chains~\cite{NadjPerge2013}, which give rise to a similar induced spin-orbit coupling and Zeeman splitting of the electronic bands. However, these earlier treatments neglect the dynamical effects of the spin-electron interaction, which here gives rise to an intrinsic triplet superconductivity.


The electron-magnon interaction is found by expressing the spin-electron coupling (the last term in Eq.~\ref{eq:hamiltonian}) in terms of the magnon and band electron operators. Due to the non-collinear structure of the spin spiral, the coupling has a non-trivial spin structure (see Fig.~\ref{fig:setup}c), and in general gives rise to magnon-mediated scattering between electrons with arbitrary spin projections. An effective electron-electron interaction is obtained by integrating out the magnons within a finite temperature functional integral formulation~\cite{Supplemental}. The effective interaction $U_{\tau_1\tau_2}^{\tau_3\tau_4}(k,k')$ implicitly depends on the magnetic structure via the spin-electron coupling $g$, the spiral momentum $q$, and the transformation matrices used to diagonalize the magnon and electron sub-systems. In the static limit the superconducting gap $\Delta_{k}^{\tau\tau'}$ is determined via the linearized gap equation
\begin{align}\label{eq:gap}
 \Delta_{k}^{\tau_1\tau_2} &= \sum_{k'\tau_3\tau_4} U_{\tau_1\tau_2}^{\tau_3\tau_4}(k,k') \Big( \sum_{\tau} \frac{\tanh(\beta \xi_{k'\tau})}{2\xi_{k'\tau}} \Big) \Delta_{k'}^{\tau_3\tau_4},
\end{align}
where $\beta = 1/(k_BT)$ is the inverse temperature, and $\xi_{k} = \epsilon_{k} - \mu$. The gap equation can be viewed as an eigenvalue problem for the susceptibility matrix $\chi_{\tau_1\tau_2}^{\tau_3\tau_4}(k,k')$, implicitly defined by the right hand side of Eq.~\ref{eq:gap}, and superconductivity is signaled by the largest eigenvalue of this matrix exceeding unity~\cite{Allen1983}.


On the mean-field level the electronic Hamiltonian is of the Bogoliubov-de Gennes (BdG) form, and can be written as $ H = \sum_k \Phi_k^\dagger \mathcal{H}_k \Phi_k$ with $\mathcal{H}_k$ as a $4 \times 4$ matrix~\cite{Supplemental}. The gap $\Delta_{\tau\tau'}$ can be decomposed into four components, given by the singlet gap $\Delta_s = (\Delta_{du} - \Delta_{ud})/2$ and the triplet gaps $\Delta_{u} = \Delta_{uu}$, $\Delta_{d} = \Delta_{dd}$, and $\Delta_p = (\Delta_{du} + \Delta_{ud})/2$. The subscripts $u$ and $d$ refer to the band index $\tau$, which can be viewed as an effective spin index, where $u$ ($d$) denotes the upper (lower) band or effective spin up (down). Due to the fermionic nature of the electronic operators, the singlet and triplet gaps are even and odd functions of $k$, respectively.

We begin with analyzing how the symmetry of the dominant superconducting gap depends on the spiral momentum $q$, the spin-electron coupling $g$, and the chemical potential $\mu$. For parameters such that the energy difference $\epsilon_u(k_{Fu}) - \epsilon_d(k_{Fd}) \gg \Omega$, with $k_{F\tau}$ the Fermi momentum in band $\tau$, and $\Omega$ a typical magnon energy, Cooper pairs will predominantly form from states within the same band. In this regime, the gap function becomes diagonal in the band indexes. At temperatures below $10$~K, the pairing interaction is highly restricted to the Fermi surface, which for a 1-D system consists of the points $\pm k_{F\tau}$. Restricting the gap equation to the Fermi surface, and neglecting contributions from inter-band scattering, Eq.~\ref{eq:gap} reduces to a $2\times 2$ matrix problem and can be solved analytically~\cite{Supplemental}. The eigenvalues of the susceptibility matrix are $\lambda_s = \chi(k_F,k_F) + \chi(k_F,-k_F)$ and $\lambda_p = \chi(k_F,k_F) - \chi(k_F,-k_F)$, for singlet and triplet pairing, respectively. These solutions show that the sign of the term $\chi(k_F,-k_F)$, accounting for scattering between the points $k_F$ and $-k_F$, determines the symmetry of the dominant gap. Since only triplet pairing is consistent with the symmetry of intra-band Cooper pairs, a consistent superconducting solution requires $\lambda_p > \lambda_s$. When $\epsilon_u(k_{Fu}) - \epsilon_d(k_{Fd}) \lesssim \Omega$, inter-band scattering becomes important, and the dominant superconducting gap is expected to have $s$-wave symmetry.

This qualitative analysis is in good agreement with the numerical solution of Eq.~\ref{eq:gap}, showing that the dominant superconducting gap has triplet symmetry over a large fraction of the phase diagram (see Fig.~\ref{fig:parameters}c). Only in regions where $\epsilon_u(k_{Fu}) - \epsilon_d(k_{Fd}) \lesssim \Omega$, such as close to the ferromagnetic limit $q = 0$, or when both $k_{F\tau}$ are close to the band edges (around $k_{F\tau} = 0$), does the dominant gap have singlet symmetry. In the strict ferromagnetic limit, this result can be established analytically~\cite{Supplemental}, by noting that ferromagnetic magnons only scatter electrons with opposite spins. That such a large portion of the phase diagram is dominated by triplet pairing follows from the fact that singlet pairing requires inter-band scattering. For $k_{Fd} \neq k_{Fu}$, such scattering results in Cooper pairs with a finite center-of-mass momentum $k_{Fu} - k_{Fd} \sim q$, which strongly suppresses singlet pairing at finite $q$~\cite{Loder2010}. With increasing $q$, intra-band magnon scattering therefore quickly becomes the dominant pairing mechanism, resulting in a gap with triplet symmetry. We note that in the antiferromagnetic limit $q = \pi$, an effective gap opens between the electronic bands, and only triplet pairing is possible.


We now analyze how the magnitude of the dominant superconducting gap depends on the magnetic parameters. In the regime $\epsilon_u(k_{Fu}) - \epsilon_d(k_{Fd}) \gg \Omega$, where inter-band scattering can be neglected, the zero temperature mean-field gap is given by $\Delta = 2\Omega e^{-1/(\rho_F U)}$~\cite{Supplemental}. Here $\rho_F$ is the electronic density of states (DOS) at the Fermi level, and $U = U(k,k) - U(k,-k)$ the effective pairing interaction in the triplet sector. Since $U \sim g^2/\Omega$, the superconducting gap is expected to increase exponentially with the spin-electron coupling. To estimate $\Delta$, we note that for an approximately linear dispersion, the electronic DOS tends to the constant $\rho_0 = 1/(4\pi t)$, while close to a band edge at energy $\epsilon_0$ van Hove singularities of the form $\rho(\epsilon) = \rho_0 \sqrt{t/(\epsilon - \epsilon_0)}$ develop. Approximating $U$ with its value in the ferromagnetic limit, $U = 4g^2S/\Omega$, the dimensionless coupling strength is $\lambda_{\rm eff} = \rho_F U \approx g^2S/(\pi t \Omega)$. For the typical values $S = 1$, $t = 1$~eV and $\Omega = 10$~meV, a spin-electron coupling of $g = 100$~meV gives a gap of $\Delta \sim 1$~meV, with significant enhancements expected at the van Hove singularities. 

The magnitude of the superconducting gap is also obtained from a numerical solution of Eq.~\ref{eq:gap}, whose largest eigenvalue $\lambda$ is related to the gap by $\Delta = 2\Omega e^{-1/\lambda}$. The magnitude of $\Delta$ is shown in Fig.~\ref{fig:parameters}d, and is found to generally increase with $g$ as anticipated above. However, it also displays a clear non-monotonic behavior with $g$, which can be attributed to the variation of the electronic DOS with the magnetic parameters. In particular, a gap between the upper and lowers bands opens at $k = 0$ as a function of $g$, whose position varies with the spiral momentum $q$ (see Fig.~\ref{fig:parameters}b). For constant chemical potential, as assumed in Figs.~\ref{fig:parameters}c and \ref{fig:parameters}d, the DOS at the Fermi level therefore changes both with $g$ and $q$, and is significantly enhanced at the van Hove singularities forming at the band edges. The diagonal lines in Fig.~\ref{fig:parameters}d, emanating from $q \approx \pi/(2a)$, trace the van Hove singularities as a function of $g$. The overall magnitude of the gap, of order $\Delta \sim 1$~meV, is in line with the estimate above.


We now investigate the topological phases of the superconducting wire. In the effective single-band limit, where the chemical potential lies inside the band gap, the BdG Hamiltonian maps directly onto the Hamiltonian of the Kitaev chain~\cite{Kitaev2001}. This system is known to permit a topologically non-trivial phase with unpaired Majorana bound states appearing at each end of the quantum wire. Since the effective single-band limit of the model discussed here can be connected to the Kitaev model by continuously deforming the dispersion and gap function, its topological phase diagram is identical to the Kitaev model~\cite{Lutchyn2010,Supplemental}. It can be constructed by counting the number of intersections $N$ between a line at constant chemical potential and the dispersion, with the topological index given by $\mathbb{Z}_2 = (N/2) \mod 2$.

\begin{figure}
 \includegraphics[width=\columnwidth]{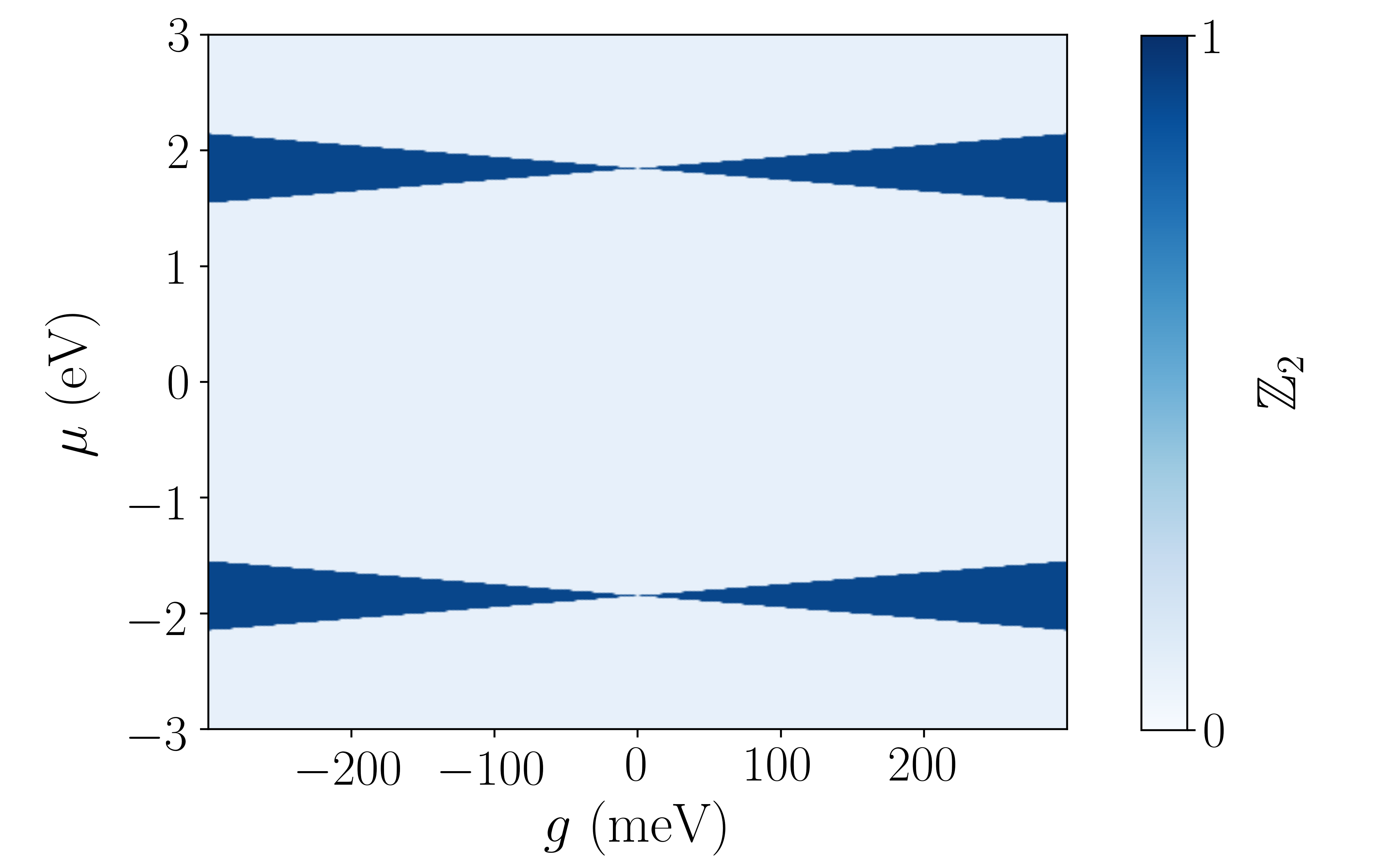}
 \caption{Topological phase diagram of magnon-mediated superconductivity as a function of spin-electron coupling $g$ and chemical potential $\mu$. The magnetic spiral has a wave length $L = 8$ and momentum $q = \pi/(4a)$. The light regions show the trivial phase, while the dark regions correspond to a topological phase. The spin length is $S = 1$ and the electronic hopping $t = 1$ eV.}
 \label{fig:invariant}
\end{figure}

To obtain the topological phase diagram of the full two-band model, we first consider the case where $\Delta_u \neq 0$ and $\Delta_d \neq 0$ but $\Delta_s = \Delta_p = 0$. In this limit, the model corresponds to two independent copies of the Kiteav model, and a $\mathbb{Z}_2$ invariant can be defined separately for each band. Denoting these invariants by $Z_d$ and $Z_u$, the $\mathbb{Z}_2$ index of the full system is given by $Z_d Z_u$. Since adding finite inter-band pairings cannot close the superconducting gap, the full two-band model is continuously connected to the $\Delta_s = \Delta_p = 0$ limit, and for $\Delta_u \neq 0$ and $\Delta_d \neq 0$ the phase diagram can be constructed by counting the number of times $N$ a line at constant chemical potential crosses the bands $\epsilon_{k\tau}$ (with $\mathbb{Z}_2 = (N/2) \mod 2$, see Fig.~\ref{fig:invariant}). For $g = 0$ the system is always in a trivial regime, while for $g \neq 0$ two regions with non-trivial topology grow out of the upper and lower band edges. The size of these regions increase linearly with the magnitude of $g$. We note the in the antiferromagnetic limit the system is always in the trivial phase, since there is no effective single-band regime in this case.


Our results demonstrate that topological superconductivity is stabilized whenever the quantum wire is in an effective single-band regime. Such a regime can be experimentally realized by placing the chemical potential inside the $k = 0$ or $k = \pm \pi/a$ band gap $\Delta_b$, induced by the coupling to the spiral. Magnon scattering further opens a superconducting gap $\Delta$, whose magnitude is shown in Fig.~\ref{fig:parameters}c and is approximately determined by the product $g^2\rho_F/\Omega$. To favor topological superconductivity, it is desirable to find a system with a large $g$ and $\rho_F$, but a small $\Omega$ (however not too small, since $\Omega$  sets the upper limit of $\Delta$).

To realize the model of Eq.~\ref{eq:hamiltonian}, several strategies are possible. The most straightforward is to consider a quantum wire in proximity to a helical magnet, as illustrated in Fig.~\ref{fig:setup}. In this case, the wire could consist either of a simple metal, or a lightly doped or gate tuned semi-conductor (such as InAs~\cite{Nilsson2018,Debbarma2022}), with a single active and largely uncorrelated band. For the helical magnet, it is preferable to use a material with a short spiral wave length ($\lambda \sim 10$~nm), low magnon energy and large spin length, such as MnGe, MnSi or NiI$_2$~\cite{Nagaosa2013,Song2022}. To maximize $g$, it is desirable to use a wire geometry where as much as possible of the wire is strongly coupled to the magnet. Estimating $g$ for a general interface is hard, but based on earlier work it is expected that $g \sim 10 - 100$~meV can be achieved~\cite{Qiao2014,Norden2019,Maeland2021,Cardoso2023}. An alternative strategy is to consider a quasi-one-dimensional multi-band system~\cite{Schlappa2012,Kennes2020}, where the magnetic order and itinerant electrons co-exist within the same material. In this case the spin-electron coupling is effectively the Hund's coupling $J$, potentially leading to large effective $g$ on the order of $1$~eV. However, since such multi-band systems are typically strongly correlated, Eq.~\ref{eq:hamiltonian} should in this case be supplemented with additional interaction terms.


We have demonstrated that magnon fluctuations of a helical magnet can mediate triplet superconductivity in a proximate quantum wire. As the induced band gap and superconducting gap both increase with the spin-electron coupling $g$, various material engineering techniques such as wire geometry optimization, microstructuring or cavity-enhanced light-matter couplings can be employed to maximize its value~\cite{Moll2018,Sentef2018,Nilsson2018,Barrigon2019,Bostrom2023}. We expect that, utilizing such strategies, an effective coupling exceeding $g \approx 50$~meV can be achieved, allowing our protocol to surpass present state-of-the-art experiments with a topological superconducting gap of the order $\Delta \sim 0.1$~meV~\cite{Microsoft2023}. Compared to the conventional scheme of generating 1-D topological superconductivity combining proximity-induced superconductivity with large Rashba spin-orbit coupling and external magnetic fields, the main advantage of our proposal is the absence of fine tuning conditions.

The present model is a simplified description of a real quantum wire, and in reality effects coming from repulsive electron-electron interactions, disorder and a more complex band structure should be included~\cite{DasSarma2023}. While the former is expected to compete with the magnon-mediated attraction, the later could result in a larger DOS at the Fermi level, thereby enhancing superconductivity. Another route to connect the present model with more realistic experimental setups is through the study of the corresponding system with open boundary conditions. In conclusion, our results identify quantum wires in proximity to non-collinear magnets as a promising platform to explore topological superconductivity in 1-D.


\begin{acknowledgments} 
We acknowledge fruitful discussions with Oladunjoye A. Awoga. FVB acknowledges funding from the Swedish Research Council (VR), and EVB acknowledges funding from the European Union's Horizon Europe research and innovation programme under the Marie Sk{\l}odowska-Curie grant agreement No 101106809. 
\end{acknowledgments}


\bibliography{references}


\end{document}